# Molecular Weight Dependent Structure and Polymer Density of the Exopolysaccharide Levan

Hundschell, C. S.[1]*, Jakob, F.[2]*, Wagemans, A. M.[3]

[1]Department of Food Technology and Food Material Science, TU Berlin;
[2]Department of Technical Microbiology, TU München;
[3]Department of Food Colloids, TU Berlin
* Corresponding author

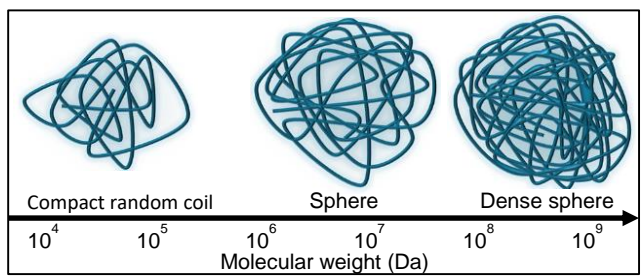

**Abstract:** Levan is a bacterial homopolysaccharide, which consists of β-2→6 linked β-(D)-fructose monomers. Because of its structural properties and its health promoting effects, levan is a promising functional ingredient for the food, cosmetic and pharma industry. The properties of levan have been reported to be linked to its molecular weight. For a better understanding of how its molecular weight determines its polymer conformation in aqueous solution, levan produced by the food grade acetic acid bacterium *Gluconobacter albidus* TMW 2.1191 was analysed over a broad molecular weight range using dynamic and static light scattering and viscometry. Levan, with low molecular weight, exhibited a compact random coil structure. As the molecular weight increased, the structure transformed into a compact non-drained sphere. The density of the sphere continued to increase with increasing molecular weight. This resulted in a negative exponent in the Mark-Houwink-Sakurada Plot. For the first time, an increase in molecular density with increasing molecular weight, as determined by a negative Mark-Houwink-Sakurada exponent, could be shown for biopolymers. Our results reveal the unique properties of high-molecular weight levan and indicate the need of further systematic studies on the structure-function relationship of levans for their targeted use in food, cosmetic and pharmaceutical applications.

# 1. Introduction

Fructans are synthesized by a variety of plants and microorganisms. In plants, short chain fructans serve as carbohydrate storage and protect against cold and dry stress (Pollock, 1986). Microorganisms produce fructans such as levan or inulin extracellularly during formation of biofilms (Velazquez-Hernandez et al., 2009). In the context of human consumption, prebiotic properties and other positive effects such as lowering cholesterol as well as antiviral and antitumoral properties have been reported for inulin and levan (Abdel-Fattah, Mahmoud, & Esawy, 2005; Belghith et al., 2012; Esawy et al., 2011; Korakli, Gänzle, & Vogel, 2002; Liu, Luo, Ye, & Zeng, 2012; Marx, Winkler, & Hartmeier, 2000).

Besides the nutritional characteristics, microbial fructans have unique techno-functional properties due to the high molecular weight fractions. Thus, the exopolysaccharide levan can be used to boost the texture and appearance in foods. For example, a positive effect on the crumb hardness and the specific volume of gluten free buckwheat bread was observed in baking experiments. As a result, a higher molecular weight was found to enhance these positive effects (Ua-Arak, Jakob, & Vogel, 2017a, 2017b). Moreover, levan can be applied as opacifying agent since strong clouding was observed in solution due to the Tyndall effect (Fig. 1), which was more pronounced at high molecular weight (Benigar et al., 2014; Huber, Stayton, Viney, & Kaplan, 1994; Jakob, 2014). These results suggest that the techno-functional properties of the exopolysaccharide levan depend strongly on its molecular weight, which can be tailored by the fermentation conditions (Ua-Arak et al., 2017a).

The enzymatic synthesis of levan is divided in several steps facilitating the formation of various molecular weights and structures. At first, the enzyme levansucrase (EC 2.4.1.10) cleaves the bond between glucose and fructose of its substrate sucrose. The energy released during hydrolysis of the glyosidic bond is used to transfer the fructose molecule to the acceptor. At the beginning of the subsequent reaction, the fructose is transferred to a sucrose molecule; but as fructosylation proceeds, one fructose molecule is bound to another producing fructooligosaccharides and finally β-2,6 linked levan which can have β-2,1 linked branches (Ozimek, Kralj, van der Maarel, & Dijhuizen, 2006). The molecular weight of levan was reported to range from $10^5$ to $10^8$ Da (Srikanth, Reddy, Siddartha, Ramaiah, & Uppuluri, 2015; Tomulescu et al., 2016). However, in previous experiments levan could also be formed with molecular weights of up to more than $2*10^9$ Da (Jakob et al., 2013).

Literature discussing various physical properties of levan reveal varying results. The radius of gyration was found to be between 20 nm and several hundred nm (Arvidson, Rinehart, & Gadala-Maria, 2006; Jakob et al., 2013; Kasapis, Morris, Gross, & Rudolph, 1994; Runyon et al., 2014), while the hydrodynamic radius ranged between 2.5 nm and 151 nm (Arvidson et al., 2006; Runyon et al., 2014). The unusual low intrinsic viscosity of levan (7 – 45 mL/g) indicated a compact and spherical molecular structure (Bae, Oh, Lee, Yoo, & Lee, 2008; Bahary & Stivala, 1978; Benigar et al., 2014; Ehrlich et al., 1975; S. S. Stivala & Zweig, 1981; Salvatore S. Stivala, Bahary, Long, Ehrlich, & Newbrun, 1975). Studies on the viscosity of levan. indicated Newtonian behaviour at concentrations up to 20 to 30 percent (Arvidson, Rinehart, and Gadala-Maria (2006) and Kasapis et al. (1994)). At higher concentrations shear-thinning behaviour was observed. In contrast, the transition point from Newtonian to shear-thinning behaviour ranged from 1 to 4 percent for levans of three different microbial origins according to Benigar et al. (2014).

In the above mentioned study, the different ranges for several levan properties can be explained by the microbiological origin, molecular weight and structural properties that differed. Moreover, the dispersity of the levan samples was not always characterized, which may have a large effect on the measured properties. In addition, to best of our knowledge no study correlated the molecular weight dependence of levan properties, especially in the molecular weight range above $10^7$ Da, over a broad molecular weight range using one type of levan produced by a specific microorganism and its secreted levansucrase.

Therefore, we aimed to characterize the structural properties of *Gluconobacter (G.) albidus* levan fractions with a low dispersity in molecular weight, in the range from $10^4$ Da to $10^9$ Da, using dynamic light scattering, multi angle light scattering and viscometry. Doing so, we correlated molecular weight, intrinsic viscosity, hydrodynamic radius, radius of gyration and geometric radius in order to investigate the change in molecular structure as a function of molecular weight.

We built upon work from previous studies, in which the microbial levan production was established using *G. albidus* (TMW 2.1191). The gene sequence responsible for the formation of the enzyme was identified by PCR and the enzyme was found to belong to the family 68 of glycoside transferases and referred to as levansucrase (Jakob, 2014; Jakob, Meißner, & Vogel, 2012). The molecular weight and the amount of the linear, β-2,6 linked levan (Fig. 2) produced by *G. albidus* can be controlled by the fermentation time and the pH (Jakob, 2014; Ua-Arak, Jakob, & Vogel, 2016; Ua-Arak et al., 2017a, 2017b)

## 2. Theoretical Background

### 2.1 Photometry

Photometry is a standard analysis that uses the extinction of light E to determine the concentration of a solute, as described by the Beer-Lambert law.

$$E = log\frac{I_0}{I_1} = \varepsilon cd \tag{1},$$

where $I_0$ and $I_1$ are the intensities of the radiated and the transmitted light, respectively, c is the concentration, d is the path length of the vessel and $\varepsilon$ is the specific extinction coefficient. The light can be attenuated by reflection, scattering, diffraction or absorption. At a wavelength of 200 nm, the glyosidic bonds of levan have their adsorption maximum. Above this wavelength, the extinction, and thus the extinction coefficient, of high molecular weight levan solutions are mainly originating from light scattering. Therefore, the extinction coefficient can also be considered as a measure of the turbidity of levan solutions. In addition to the quantities given in Beer-Lambert law, the extinction coefficient depends on the colloid size and the difference in refractive index between colloid and solvent.

### 2.2 Viscometry (Intrinsic Viscosity and Huggins Coefficient)

The reduced viscosity $\eta_{red}$ can be deduced from the solvent $\eta_0$ and solution viscosity $\eta_s$ and describes the contribution of a dissolved polymer to the total viscosity of the solution. In nonelectrolytic polymer solutions, it can be represented by a series in powers of concentration c (Huggins, 1942):

$$\eta_{red} = \frac{\eta_s - \eta_0}{\eta_0 c} = A_0 + A_1 c + A_2 c^2 + A_3 c^3 + \ldots \tag{2}$$

In dilute solutions, the $c^2$ and higher power terms are negligible. This gives the well-known Huggins equation where $A_0$ is the intrinsic viscosity $[\eta]$, and $A_1$ is equal to $k_H[\eta]^2$ with the Huggins coefficient $k_H$ (Huggins, 1942).

$$\eta_{red} = \frac{\eta_s - \eta_0}{\eta_0 c} = [\eta] + k_H[\eta]^2 c \tag{3}$$

The intrinsic viscosity is a hydrodynamic parameter that reflects the volume occupied by a given polymer mass. It depends on the size and conformation of the polymer chain. The intrinsic viscosity corresponds to the y-intercept of the Huggins equation and is determined by extrapolating $\eta_{red}$ to zero concentration. The Huggins coefficient is related to the slope of the Huggins equation. In the literature, $k_H$ is used to describe the nature of interactions or the affinity in a polymer-solvent system. Therefore, it is used as a measure of solvent quality: the better the solvent quality the lower the $k_H$ value. In a good solvent, the polymer-solvent interactions dominate the system and $k_H$ attains a value around 0.3. At theta conditions, polymer-polymer and polymer-solvent interactions are equal and $k_H$ is about 0.5 – 0.7. In most cases, $k_H$ is independent of the molecular weight, however $k_H$ was reported to be sensitive to molecular association (Sakai, 1968; Schoff, 1999). A molecular weight dependency can occur for the $k_H$ of polymers that readily associate via ionic or polar interactions or through hydrogen bonding (Dort, 1988). An overview of the Huggins constant for biopolymers is provided by the work of Pamies et al. (2008).

## 2.3 Dynamic Light Scattering (DLS)

A DLS instrument measures time-dependent intensity fluctuations of scattered light originating from the (diffusive) motion of scattering particles (macromolecules) in the solution. The superposition of the scattered light leads to constructive or destructive interferences and results in variations of the light scattering intensity with time. The diffusion coefficient (D) can be determined from these intensity fluctuations using an autocorrelation function. By means of the Stokes-Einstein equation, D can be used to determine the hydrodynamic radius $R_{H,DLS}$ (Teraoka, 2002):

$$R_{H,DLS} = \frac{kT}{6\pi\eta_0 D} \tag{4}$$

Where k is the Bolzmann constant and T is the temperature. Assuming the polymer has a spherical shape, the hydrodynamic radius can also be calculated from the intrinsic viscosity and molecular weight M.

$$R_{H,visc} = \left(\frac{3[\eta_0]M}{10\pi N_a}\right)^{\frac{1}{3}} \tag{5}$$

Equation 5 is derived from the viscosity law according to Einstein. Therefore, the hydrodynamic radius is also called the Einstein radius ($R_{H,visc}$) or the viscosity radius (Armstrong, Wenby, Meiselman, & Fisher, 2004).

### 2.4 Multi Angle Laser Light Scattering

Multi angle laser light scattering (MALLS) allows the determination of molecular weight, the radius of gyration $R_G$ and the second osmotic virial coefficient $A_2$. The ZIMM notation based on the Raylight-Debey-Gans light-scattering model offers the basis for the evaluation of light scattering data (Wyatt, 1993).

$$\frac{K^*c}{R_\Theta} = \frac{1}{MP_\Theta} + 2A_2c \tag{6}$$

where $R_\Theta$ is the Rayleigh ratio as a function of scattering angle $\Theta$. The angular dependence of the scattered light $P_\Theta$ is related to the radius of gyration. The first order expansion of $P_\Theta$ and the optical parameter $K^*$ are given by (Wyatt, 1993):

$$P_\Theta = 1 - \frac{16\pi^2 n_0^2}{3\lambda_0^2} R_G^2 \sin^2\left(\frac{\Theta}{2}\right) \tag{7}$$

$$K^* = \frac{4\pi n_0^2}{N_A \lambda_0^4}\left(\frac{dn}{dc}\right)^2 \tag{8}$$

where $n_0$ is the refractive index of the solvent, $\lambda_0$ is the wavelength in vacuum of the incident light and $dn/dc$ is the refractive index increment. To analyse the scattering data, $K^*c/R_\Theta$ is plotted against $\sin^2(\Theta/2)$ (ZIMM-Plot). From the slope of the linear extrapolation to the zero angle one obtains the radius of gyration and from its intercept with the y-axis the weight average molecular mass.

If particles are significantly smaller than the wavelength of the light, they scatter the light isotopically and $K^*c/R_\Theta$ is linear over the entire angular range. For particles whose size is in the range of the wavelength or above, non-linear scattering occurs. The curvature of these functions depends on the particle size and the particle shape and is reflected in the angular dependence of $P_\Theta$. In order to account for this dependency when extrapolating to zero angle, it may be necessary to adjust the theoretical expressions accordingly. Therefore, in addition to the ZIMM-model, $\sin^2(\Theta/2)$ is also frequently fitted against $R_\Theta/K^*c$ (Debey-model) or $(R_\Theta/K^*c)^{1/2}$ (Berry-model) to determine the molecular weight or the radius of gyration. In addition to these models, which are independent of the polymer conformation, models that presuppose a polymer structure have been developed. The following equation describes the angular dependence of scattered light for a model assuming a spherical confirmation (van de Hulst, 1957).

$$P_\Theta = \frac{3}{u^3}(\sin u - u \cos u), \text{ with} \qquad u = 4\pi \left(\frac{R_{geo}}{\lambda}\right) \sin\left(\frac{\Theta}{2}\right) \qquad (9)$$

In contrast to the conformation-independent models, the spherical model determines the geometric radius $R_{geo}$ instead of the radius of gyration. The geometric radius describes the actual spatial dimension of a spherical molecule while the radius of gyration represents the mean distance of the individual mass segments of a polymer to its centre of gravity.

### 2.5 Molecular Conformation and Shape of Macromolecules

The various measures describing molecular size obtained by the methods discussed above, can be related to each other to determine the conformation and shape of a macromolecule.

Using MALLS, the weight average molecular weight ($M_w$) and the number average molecular weight ($M_n$) can be determined. The quotient of $M_w/M_n$ defines the dispersity index PDI. The PDI is a measure of the width of a molecular weight distribution. Monodisperse samples have a PDI of 1. Because the distribution of polysaccharides is usually not monodisperse, the PDI is > 1 and increases as the molecular weight distribution broadens.

The hydrodynamic radius $R_H$ of a polymer is an apparent measure and corresponds to the radius of an equivalent hard sphere with the same diffusion properties as the observed macromolecule. Therefore, the value of the hydrodynamic radius and the geometric radius is essentially identical for compact, spherical macromolecules. For polymer conformations that differ from the compact spherical shape, the values of hydrodynamic and geometric radius are different (Pecora, 2000). The radius of gyration $R_G$ specifies the mean distance between the individual polymer segments of a molecule and its centre of mass. The ratio of the radius of gyration and hydrodynamic radius provides information about the investigated polymer's shape. In theory, $R_G/R_H$ for a compact sphere is 0.775 while less compact polymers exhibit a higher $R_G/R_H$. According to Nilsson (2013), $R_G/R_H$ for linear random coils under theta conditions is 1.5 and in good solvents is 1.78. A listing of $R_G/R_H$ values for some polysaccharides can be found in the publication by Nilsson (2013). A similar quotient can be defined for the hydrodynamic and the geometric Radius ($R_H/R_{geo}$). In a non-drained sphere, the hydrodynamic radius and the geometric radius are equal and therefore $R_H/R_{geo}$ is 1 (Pecora, 2000). In a less compact polymer, the solvent can permeate through the coil resulting in different diffusion properties including a decrease in the hydrodynamic radius which makes it virtually smaller than the geometric radius. Considering this behaviour, $R_H/R_{geo}$ decreases with increasing polymer density.

The conformation of polymers can be analysed by linking the radius of gyration with the molecular weight (Nilsson, 2013).

$$R_G = k_G M^{\nu_G} \qquad \qquad 10$$

The constant $k_G$ depends on the polymer and solvent. The hydrodynamic coefficient $\nu_G$ depends on the spatial structure and density of the macromolecules. For a compact sphere $\nu_G$ is 0.33, for a random coil 0.5 – 0.6 and for a rod 1 (Nilsson, 2013). A smaller $\nu_G$ can be interpreted as to reflect a more compact (denser) molecule structure. The Mark-Houwink-Sakurada relation (Eq. 11) is the equivalent to equation 10 (Teraoka, 2002).

$$[\eta] = k_\eta M^\alpha \qquad \qquad 11$$

Instead of the radius of gyration, the intrinsic viscosity is plotted against the molecular weight. The constant $k_\eta$, depends, like $k_G$, on the polymer-solvent system. The Mark-Houwink-Sakurada exponent α gives information about the molecule structure. A compact sphere has an α of 0. Under theta conditions, α for a linear random coil is 0.5 and for a rod it 1 (Teraoka, 2002). Both exponents, $\nu_G$ and α, are linked by the following equation (Öttinger, 1996).

$$\alpha = 3\nu_G - 1 \qquad \qquad 12$$

### 3. Material & Methods

#### 3.1 Cultivation and Levan Production

The acetic acid bacterium *Gluconobacter* (*G.*) *albidus* TMW 2.1191, isolated from water kefir (Gulitz, Stadie, Wenning, Ehrmann, & Vogel, 2011), was cultivated in sodium gluconate medium (NaG) containing 20 g/L sodium gluconate, 3 g/L yeast extract, 2 g/L peptone from casein, 3 g/L glycerin, 10 g/L mannitol and 3 g/L glucose.

For fermentative levan production, mannitol and glucose were replaced by 80 g/L of sucrose. For agar plates, 20 g/L of agar was added.

To obtain a preculture, a single colony of *G. albidus* was transferred into a 500 ml Erlenmeyer flask containing 50 ml of NaG medium and cultivated overnight at 30 °C and 200 rpm in a rotary shaker to an optical density of approximately 2.5.

### 3.1.1 Fermentative Levan Production

After pre-cultivation, the microorganisms were centrifuged for 5 min at 5000 g and re-suspended in 5 ml of fresh sucrose-containing NaG medium. 1 ml of this suspension was used to inoculate 200 ml of sucrose-containing NaG medium in 2 L Erlenmeyer flasks. After growth for 48 hours at 30 °C and 200 rpm, the cells were removed by centrifugation (10,000 g, 15 min). The levan in the supernatant was precipitated with twice the volume of ethanol and stored overnight at 4 °C. The subsequent purification of the levan was performed according to Jakob et al. (2013). Levan, which was produced in cell-containing medium, is abbreviated with LevF in the following part of the manuscript (F in reference to the fermentative levan production process in presence of metabolic active cells).

### 3.1.2 Enzymatic Levan Production

The treatment of the preculture and inoculation of the working culture was carried out using sucrose-free NaG medium, as described in 3.1.1. After a cultivation of 24 hours at 30 °C and 200 rpm, the cells were centrifuged for 10 minutes and re-suspended in 200 ml of 0.1 M sodium acetate buffer containing 0.1 M sucrose in 2 L Erlenmeyer flasks according to Jakob (2014). The buffer was adjusted to pH 4 or pH 5 to produce levans with different molecular weights. To isolate the enzyme, the cell-containing buffer was incubated for 3 h at 30 °C and 200 rpm in a rotary shaker. Subsequently, the cells were centrifuged and 200 mL of 0.1 M sodium acetate buffer with 0.7 M sucrose was added to the enzyme-containing buffer. The subsequent levan production was performed at 30 °C for 24 h. The separation of the levan from the buffer and the purification were carried out as described in 3.1.1. Levan produced using levansucrase containing, cell-free supernatants is abbreviated to Lev4 and Lev5 (4 and 5 in reference to the pH of the used sodium acetate buffer during cell-free levan production).

### 3.1.3 Gradual Ethanol Precipitation

For each levan, 4 to 7 fractions were obtained by gradual ethanol precipitation. 10 g of each levan were dissolved in 500 mL of distilled water and stirred overnight at 4 ° C. Afterwards, ethanol was added slowly while continuously stirring until a change in turbidity indicated that the levan had precipitated. Subsequently, the precipitated levan was separated from the solution by centrifugation (10000 g, 30 min, 4 °C). This procedure was repeated with the remaining supernatant until no further precipitation was visible. Ethanol concentrations and sample amounts are listed in Table 1. After centrifugation, the precipitate was washed with ethanol, the remaining ethanol was evaporated at room temperature, and the samples were freeze-dried.

Table 1: Ethanol concentrations and levan amounts of the fractionation process

| Levan | Fraction | Sample name | Ethanol (vol %) | Mass (g) |
|---|---|---|---|---|
| LevF | 1 | LevF F1 | 55.58 | 1,15 |
|  | 2 | LevF F2 | 56.88 | 1,84 |
|  | 3 | LevF F3 | 60.69 | 1,92 |
|  | 4 | LevF F4 | 64.00 | 1,09 |
|  | 5 | LevF F5 | 70.32 | 1,38 |
| Lev4 | 1 | Lev4 F1 | 53.88 | 0,67 |
|  | 2 | Lev4 F2 | 54.88 | 1,10 |
|  | 3 | Lev4 F3 | 55.29 | 1,19 |
|  | 4 | Lev4 F4 | 55.30 | 1,29 |
|  | 5 | Lev4 F5 | 55.62 | 1,66 |
|  | 6 | Lev4 F6 | 55.97 | 0,98 |
|  | 7 | Lev4 F7 | 66.50 | 1,46 |
| Lev5 | 1 | Lev5 F1 | 52.04 | 1,55 |
|  | 2 | Lev5 F2 | 53.13 | 1,58 |
|  | 3 | Lev5 F3 | 54.15 | 4,00 |
|  | 4 | Lev5 F1 | 55.31 | 1,38 |

### 3.2 Analytics

#### 3.2.1 Sample Preparation

For all experiments, levan was dissolved in distilled water on a magnetic stir plate. The samples were stored overnight in the refrigerator to ensure complete hydration of the polymers. All analyses described below were performed in triplicate at 20 °C unless otherwise noted.

#### 3.2.2 Photometry

The specific extinction coefficient of levan was determined at 400 nm in a photometer (Helios Omega UV-vis spectrophotometer, Thermo Fisher Scientific Germany BV & Co KG, Germany). A standard curve with eight different concentrations was generated. Low molecular weight levans (LevF, LevF F2 - F5) were measured at concentrations between 20 and 50 g/L, medium molecular weight levans (LevF F1, Lev4 F7) at 3 - 5 g/L, and the higher molecular weight levans (Lev4, Lev4 F1 – F6, Lev5, Lev5 F1 – F4) between 1 and 2 g/L.

### 3.2.3 Viscometry

The intrinsic viscosity [η] was determined utilizing a rolling ball micro viscometer (LOVIS 2000M, Anton Paar GmbH, Germany). A glass capillary with a radius of 1.59 mm and a steel ball with a radius of 1.5 mm at an angle of 50 ° (shear rate 420 – 580 $s^{-1}$) were used. Viscosity was measured in triplicate at six concentrations in the range of 2.0 to 4.5 g/L. To determine the density of the solutions a bending vibrator (DMA 38, Anton Paar GmbH, Germany) was used.

### 3.2.4 Dynamic Light Scattering

The hydrodynamic radius $R_H$ of the levan fractions were measured with a ZetaSizer Nano ZS (Malvern Instruments, UK) utilizing dynamic light scattering. Each sample was measured three times with automatic measurement duration at an angle of 173°. The solvent refractive index was 1.33, and the viscosity was 1.0031 mPa s.. Low molecular levans (LevF, LevF F2 - F5) were measured at a concentration of 2.0 g/L while higher molecular weight levans (LevF F1, Lev4, Lev4 F1 – F7, Lev5, Lev5 F1 – F4) were measured at 0.1 g/l. The volume distribution and the mean of the volume distribution $R_H$ were determined. To convert the intensity distribution into the volume distribution, a refractive index of 1,65 was determined for levan. Therefore, a Horiba particle sizer (LA-950, Horiba Jobin Yvon GmbH, Bernsheim, Germany) and the Method expert option within the LA-950 software was used as described by Krzeminski et al. (2014). All data analysis was performed using the instrument software.

### 3.2.5 Multi Angle Laser Light Scattering

Levan fractions and the unfractionated levan were separated according to polymer size with asymmetric flow field flow fractionation (aF4) (Wyatt Technology, Germany) and analysed with multi-angle laser light scattering (MALLS) (Dawn Heleos II, Wyatt Technology, Germany). UV (Dionex Ultimate 3000, Thermo Fisher Scientific, USA) and RI (Refractomax 521, Thermo Fisher Scientific, USA) detection were used for concentration determination. The injected volume was 100 µL, and the polysaccharide concentration was 0.1 g/L – 1.0 g/L. For separation, the method of Jakob et al. (2013) was used with modification. An injection flow rate of 0.2 mL/min and a constant elution flow rate of 1 mL/min were used. The crossflow was reduced from 3 mL/min to 0.1 mL/min within the first 10 min of elution. Subsequently, the crossflow was kept constant at 0.1 mL/min for 15 min before being set to 0 mL/min. The separation was performed on 10 kDa regenerated cellulose membranes (Superon GmbH Germany) using a 50 mM $NaNO_3$ solution as eluent. Data are representative of two measurements.

**Concentration determination in aF4-MALLS experiments:** For low molecular weight levans (LevF, LevF F1 - F5, Lev4 F7), a RI detector was used for concentration determination. To compensate RI signal influencing pressure fluctuations, caused by the flow profile of the aF4 separation, the baseline of a water run was subtracted from the measurement signal. For levans with larger molecular sizes a UV detector at 400 nm was used for concentration determination. At this wavelength, the extinction is caused by light scattering and depends on the size of the levan. Therefore, the mean geometric radii were determined in the particle mode (no concentration signals needed for distribution analysis) and correlated with the extinction coefficients. This correlation was used to determine the concentration from the geometric radius, and the UV-signal for each volume slice during the aF4-MALLS UV measurement.

**Analysis of MALLS data:** ASTRA 6 software (Wyatt Technology, Germany) was used for data analysis. The molecular weight and the geometric radius were determined via a sphere model. For the determination of the radius of gyration, the Zimm, Debye and the Berry models were compared. After the evaluation of these different models, the MALLS data was analysed through a Debye model with a third order polynomial fit. All radii were determined in the concentration-independent particle mode. For molecular weight calculations, a refractive index increment (dn/dc) of 0.146 ml/g (50 mM $NaNO_3$) was used for *G. albidus* levan according to Ua-Arak et al. (2017). In all calculations, the concentration is assumed to be sufficiently low to neglect the second virial coefficient and higher terms. The samples should be sufficiently diluted by Af4 separation to prevent errors from neglecting the virial coefficients (Andersson, Wittgren, & Wahlund, 2003).

## 4. Results and Discussion

### 4.1 Levan Fractionation

Fractional precipitation with ethanol separated levans by their size. From the produced levan samples (3.1.1, 3.1.2), a total of 16 fractions with varying polymer size were obtained. The high molecular weight levans precipitated at ethanol concentrations of 52 – 56 vol.%, while the low molecular weight levans precipitated at significantly higher ethanol concentrations (up to 70 vol.%). Fractionation reduced the dispersity of most samples. The dispersity of the fractions was governed by the volume of ethanol added during the precipitation step and the initial size distribution of the unfractionated polymer. The cumulative distribution of the hydrodynamic radius of unfractionated levan and its fractions is shown in Figure 3. Due to the fractional precipitation with ethanol, a reduction of the width of the size distribution could be achieved. The dispersity index, calculated from aF4-MALLS data, for most fractions was found between 1.2 and 2.3 and is smaller than the PDI of the unfractionated polysaccharide. Only the last fractions of Lev4 showed a higher PDI value. Also, the first two fractions of LevF indicated a slightly broader size distribution compared to the other fractions. The dispersity index, the molecular weight and the measured radii of levan samples and its fractions are listed in the supplementary data (Tab. S1).

### 4.2 AF4 MALLS Analysis - Determination of Geometric Radius, Radius of Gyration and Molecular Weight

Depending on the molecular size of the levan, UV or RI detection was used for concentration determination in the aF4-MALLS system. The RI detector measures the difference in refractive index between solvent and solution. Therefore, it is suitable for all substances that differ in refractive index from the solvent; however, it is not as sensitive as other concentration detectors. Moreover, the refractive index is temperature and pressure dependent (Munk, 1993). For the quality of the aF4 separation and accurate concentration determination, the polymer concentration has opposite effects. The higher the concentration the worse the aF4 separation but the better the concentration signal. The low molecular weight levan could be effectively separated at higher polymer concentrations than the high molecular weight levan. Therefore, the RI detector was suitable for the concentration determination of the low molecular weight levan. The UV detector was used for high molecular weight levans because they effectively scatter light and cause a strong UV signal at low levan concentrations. Therefore, the UV detector was superior to the RI detector in terms of high molecular weight levan concentration sensitivity. In addition, no baseline subtraction was needed to compensate for the pressure dependence as applied for the RI detector. The extinction coefficient changes with molecular size; therefore, a fit (Fig. S8 supplementary data ; Eq. 13) between the geometric radius and the extinction coefficient was applied to avoid errors due to molecular size dependence:

$$\varepsilon = 3.17 \ast 10^{-6} R_{geo}^{2,23} \qquad \qquad 13$$

Equation 13 enables to assess the polymer concentration from the geometric radius and the UV detector signal. Figure S9 (supplementary data) shows the concentration curve during the aF4-MALLS measurement for Lev4 F5 and Lev4 F6. For both fractions, a suitable UV and RI signal could be obtained. After the correction for the size-dependent extinction coefficient, the concentration curves and molecular weight distributions (Fig. S10 supplementary data) from the UV and RI detection were found to be similar.

Different models can be used to determine the radius and the molecular weight from MALLS experiments. The lower limit for the determination of radii is 10 – 15 nm because below this limit the scattering of particles is not angle dependent. In respect of polymer size and conformation, the quality of the fit differed from model to model. The sphere model delivered the best fit over the entire molecular weight range for all levan fractions. Therefore, this model was used to determine the molecular weight and the geometric radius. The radius of gyration was determined using the Debye model with a polynomial third order fit. According to Andersson, Wittgren & Wahlund (2003) and Baborowski & Friese (2005), this model is well suited for compact spherical colloids with a geometrical radius of up to ~170 nm. In addition to the sphere and the Debye model, the MALLS data were also evaluated with the Zimm and the Berry model. The radius of gyration determined by these two models was, up to a geometric radius of 70 nm, comparable to those of the Debye model. Above 70 nm, the coefficient of determination of the $K^*c/R_\Theta$ against $\sin^2(\Theta/2)$ fit of the Zimm model decreased with increasing polymer size. Assuming the non-drained sphere relationship $R_H = R_{geo} = 0.775 R_G$ for a compact spherical levan, above 50 nm, the radius of gyration determined by the Zimm model tends to be overestimated (Pecora, 2000; Runyon et al., 2014). In the Berry model, $(R_\Theta/K^*c)^{1/2}$ is plotted against $\sin^2(\Theta/2)$. The coefficient of determination of this fit was comparable to the error of the sphere model; however, the radii of gyration were also overestimated with this model above a geometric radius of 70 nm. The matching trends of the two models were also found for spherical colloid by Andersson, Wittgren & Wahlund (2003) and Baborowski & Friese (2005) s.

### 4.3 Size and Conformation

The conformation of levan has been discussed in earlier studies (Bahary & Stivala, 1975; Benigar et al., 2014; Ehrlich et al., 1975; Jakob et al., 2013; Runyon et al., 2014; S. S. Stivala & Zweig, 1981; Salvatore S. Stivala et al., 1975). In most cases, only a limited range of molecular weights was investigated or levan from various microorganisms was analysed. In our study, the structure of levan from *G. albidus* was systematically analysed over a broad range of molecular weights ($10^4 – 10^9$ Da) by MALLS, DLS and viscometry.

Assuming a spherical molecular shape, the hydrodynamic radius can be determined by DLS or Equation 5. The comparison of $R_{H,DLS}$ and $R_{H,visc}$ (Fig. 4) allows to evaluate the data on their reliability since the hydrodynamic radius is determined either based on the particle motion detected by the DLS method or the intrinsic viscosity determined utilizing the viscometry, which are two independent experimental techniques. However, it should be considered that a high dispersity negatively affects the accuracy when determining the effective hydrodynamic radiusfrom DLS results. For the levan from *G. albidus*, the values for the hydrodynamic radius from DLS and viscometry were found to be almost equal. In the molecular weight range between $10^7$ Da and $5*10^8$ Da (PDI: 1.2 – 2.5), the radii differed by less than 5 percent suggesting a spherical molecular shape. Wolff et al. (2000) came to the same conclusion studying a high molecular weight inulin with a compact molecular structure. In their study, the difference between the hydrodynamic radius from DLS (108 nm) and viscometry (109 nm) of the β-2,1 linked fructan was only 1 nm. The deviation between both radii of the *G. albidus* levan were larger below $10^7$ Da (PDI: 2.0 – 8.1). This can be explained by a less compact molecular structure and an increased PDI for these fractions. The largest deviations between hydrodynamic radius from DLS and viscometry were found in the samples with the highest PDI (LevF, PDI: 19.9 and Lev4 F7, PDI 31.1).

The shape of a polymer can be estimated from the quotient of the radius of gyration and hydrodynamic radius from DLS ($R_G/R_{H,DLS}$) or radius of gyration and hydrodynamic radius from viscometry ($R_G/R_{H,visc}$). By replacing the radius of gyration with the geometric radius, $R_{geo}/R_{H,DLS}$ and $R_{geo}/R_{H,visc}$ is obtained. These quotients allow differentiation between compact spherical and the less dense random coil structures. The theoretical value for a compact sphere is 1 for $R_{geo}/R_H$ and 0.775 for $R_G/R_H$. In Figure 5, the quotients are plotted as functions of the molecular weight. For the calculation of both quotients, the hydrodynamic radius from DLS and from viscometry were used. Above $10^7$ Da ($R_H$ = 33 nm), an $R_{geo}/R_{H,DLS}$ value between 0.95 and 1.07 and an $R_{geo}/R_{H,visc}$ value between 0.98 and 1.10 indicate a spherical shape . Similarly, above $10^7$ Da, $R_G/R_{H,DLS}$ was between 0.75 and 1.03 and $R_G/R_{H,visc}$ ranged from 0.69 to 1.03 thus falling within the theoretical range for a compact spherical polymer. For less dense, hyper branched polymers, such as amylopectin-rich starch, $R_G/R_H$ was found to range from 0.99 - 1.33 (Roger & Colonna, 1999). For molecular weights above $10^8$ Da a slight decrease of $R_G/R_H$ with increasing molecular weight was observed (Fig. 5 A). This decrease was probably not related to the structure of levan but might be caused by a lack of data points at low scattering angles. For $R_{geo}$ larger than 150 nm, the resolution of the MALLS at small scattering angles is no longer sufficient to accurately determine $R_g$ with the Debye model. In the publication of Baborowski & Friese (2005), the radius of gyration of a spherical colloid above a geometrical radius of 170 nm could not be determined accurately using the Debye model with a third-order fit. In their study, the model underestimated the gyration radii. This can also be assumed for high molecular weight levan because the molecular weight should not affect the ratio of the gyration radius and hydrodynamic radius assuming a compact spherical molecular structure. This was also confirmed in our study since no change in the $R_{geo}/R_H$ value was observed for this size range. Moreover, for the levan of *G. albidus* the sphere model, which determines the geometric radius, was also the most appropriate model over the entire molecular weight range studied. For molecular weights of $10^7$ Da and less, the values of $R_G/R_{H,DLS}$ and $R_G/R_{H,visc}$ of *G. albidus* levans were between 1.48 – 2.29 and 1.22 – 1.93, respectively, and tended to increase with decreasing molecular weight. This suggests an expansion of the polymer structure with decreasing molecular weight. For a linear random coil polymer at theta conditions, a value of 1.5 is expected for $R_G/R_H$. In a good solvent $R_G/R_H$ is 1.78 (Nilsson, 2013). Runyon et al. (2014) also used $R_G/R_H$ for the structural characterization of a levan with two major size populations ranging from 2.5 to 151 nm ($R_H$). In the range of 32 nm to 129 nm, the shape factor approximately corresponded to the theoretical value of a compact sphere. Below 32 nm (33 nm in our study), their $R_G/R_H$ also increased from approximately 0.775 to 1.8 with decreasing molecular weight. Furthermore, these results demonstrated the transition from a sphere to a random coil conformation for *G. albidus* levan in the same molecular size range as observed in our study.

The hydrodynamic coefficient $v_G$ can be calculated from the slope of the double logarithmic plot of the radius of gyration as function of the molecular weight. Also the hydrodynamic radius can be used, to obtain the hydrodynamic coefficient $v_H$, which can be interpreted as $v_G$. (Btat, Kato, Katsuki, & Takahashi, 1984; Roger, Baud, & Colonna, 2001). In our study, in addition to $v_G$ the geometric radius was plotted as a function of the molecular weight to calculate the hydrodynamic coefficient ($v_{geo}$) for the levans produced from *G. albidus* because $R_{geo} = R_H$ was accurate, and the sphere model was predictive over the whole molecular weight range tested (Fig. 6). Therefore, we assumed that $v_{geo}$ can be interpreted as $v_G$. For molecular weights up to $10^8$ Da, both coefficients were largely in agreement with slightly higher values for $v_G$. Above $10^8$ Da, $v_G$ decreased more than $v_{geo}$. This can be ascribed to the limitations of the Debye model used for fitting. An insufficient amount of data points at small scattering angles for large radii of the levan could lead to an underestimation of the radius of gyration. In general, the change of $v$ with the molecular size indicates a change in the molecular conformation. In the low molecular weight range, a $v$ between 0.33 - 0.5 was found. This is true for compact random coil structures or structures between sphere and a random coil conformation. For a molecular weight around $10^7$ Da with $v$ equal to 0.33, a spherical conformation was observed. The further decrease of $v$ above a size of $10^7$ Da also indicated a spherical structure and in addition an increasing polymer density with increasing molecular weight. Slightly higher values of $v_G$ (0.43) in the molecular weight range from $18.5*10^6$ Da to $57.1*10^6$ Da could be found for a levan from *S. salivaris* (Salvatore S. Stivala et al., 1975). Jakob et al. (2013) studied the conformation of levans from four different acetic acid bacteria over a wide molecular weight range using aF4-MALLS. In the lower molecular weight range, the $v_G$ values were slightly higher than those for the *G. albidus* levans in our study. While a similar decrease in $v_G$ with increasing molecular weight was observed in both studies. In the study mentioned above, the two levans with high molecular weights produced by *Kozakia baliensis* and *Neoasaia chiangmaiensis* showed a $v_G$ less than 0.33 (Jakob et al., 2013). A similar low v was found for high molecular weight inulin (Wolff et al., 2000).

The intrinsic viscosity of *G. albidus* levan and its fractions was found between 14 mL/g and 50 mL/g. These values are unusually low for polysaccharides, but they are consistent with literature values of levan ranging from 7 mL/g to 0.45 mL/g (Bae et al., 2008; Bahary & Stivala, 1978; Benigar et al., 2014; Ehrlich et al., 1975; S. S. Stivala & Zweig, 1981; Salvatore S. Stivala et al., 1975). The low intrinsic viscosity at high molecular weights revealed the compact molecular structure of the polysaccharide. Other polysaccharides with less dense structure such as dextran or xanthan have intrinsic viscosities ranging from 100 mL/g to 50000 mL/g (Brunchi, Morariu, & Bercea, 2014; Masuelli, 2013). In addition, the exponent α in the Mark-Houwink-Sakurada plot (Fig. 7A; Eq. 11) is linked to the molecular structure. An α of 0.35 in the low molecular weight range (< $5.5*10^5$ Da) suggested a compact random coil structure for levan. In the range between $10^6$ to $10^8$ Da, the intrinsic viscosity seemed to be largely independent of the molecular weight indicating a compact spherical polymer structure. However, not enough data points were available for a detailed analysis of the Mark-Houwink-Sakurada plot in this molecular weight range. The two outliers in this range could be attributed to the high dispersity of LevF and Lev4 F7. The two domains demonstrated by two different α values for the levan in our study has been previously described. Stivala and Zweig (1981) found a random coil structure (α = 0.67) for $10^4$ Da to $8.9*10^4$ Da and a spherical structure (α = 0.05) for $2.2*10^5$ Da to $8.3*10^6$ Da. Bahary and Stivala (1978) also found two different regimes in the Mark-Houwink-Sakurada plot of acid hydrolysed levan indicating a random coil structure at low molecular weights and a spherical shape at higher molecular weights. A slightly higher α (0.17) at molecular weights from $18.5*10^6$ Da to $57.1*10^6$ Da was found by (Stivala et al., 1975). Similarly, the molecular structure of inulin was reported to change at approximately $10^5$ Da (Kitamura, Hirano, & Takeo, 1994; Wolff et al., 2000). An unusual behaviour of the *G. albidus* levan was observed at molecular weights above $10^8$ Da where the intrinsic viscosity decreased with increasing molecular weight. This resulted in a negative α (-0.12) in the Mark-Houwink-Sakurada plot. To the best of our knowledge, a negative α has not yet been reported for polysaccharides or biopolymers.

A negative α could also be calculated by equation 12, since v was found to be below 0.33 at molecular weights above $10^8$ Da (Öttinger, 1996). Therefore, a hydrodynamic coefficient smaller than 0.33 requires a negative Mark-Houwink-Sakurada exponent. Only synthetically produced, highly ordered dendric macromolecules such as polyether dendrimers show the same molecular weight dependence for the intrinsic viscosity (Fréchet, 1994; Mourey et al., 1992). In our study, the intrinsic viscosity increased with molecular weight, then reached its maximum and decreased with further increasing molecular weight. The decrease in intrinsic viscosity at high molecular weights also suggests an increasing molecular density for the spherical levan molecules with increasing molecular weight. Since the intrinsic viscosity is a measure of the hydrodynamic volume and refers to a volume per gram, the results are not in contradiction to the continuously increasing radii determined by DLS and MALLS measurements. With rising molecular density, fewer molecules are present per gram of polymer. The radius of these polymers rises with molecular weight, but the volume per gram decreases due to the rise in molecular density.

The second term of the Huggins equation (Eq. 3) provides the Huggins coefficient. This is a measure of solvent quality and can provide additional information about polymer-polymer interactions. The Huggins coefficient was found to be largely independent of the molecular weight between $10^4$ to $10^7$ Da with values of 0.75 to 0.95 indicating poor solvent properties of distilled water at 20 °C for levan (Fig 7 B). The Huggins coefficient was also located slightly above the theta value (0.5 – 0.7), which is where polymer-solvent and polymer-polymer interactions are equal. Similar values for the Huggins coefficient were found by Ehrlich et al. (1975) for fractions of a *S. salevarius* levan with a constant Huggins coefficient of approximately 1.0. The molecular weight of the unfractionated polymer was $31.5*10^6$ Da. Therefore, the fractions should be in the same molecular weight range as the *G. albidus* levan at molecular weights up to $10^7$ Da. The Huggins coefficient increased above $10^8$ Da with increasing molecular weight indicating that polymer-polymer interactions are favoured. According to Dort (1988) a molecular weight dependent Huggins constant indicates attractive interactions caused by hydrogen bonds, polar or ionic interactions between the polymers.

## 4.4 General Discussion

The results of the DLS, MALLS and viscosity measurements suggest that the *G. albidus* levan conformation is divided into three areas. At low molecular weights (< $5.5*10^5$ Da), the quotients of the radii, α and ν suggest a compact random coil structure. In the intermediate region around $10^7$ Da, a ν of approximately 0.33 and a largely molecular weight independent intrinsic viscosity suggest a compact spherical structure. This spherical shaped structure is retained at higher molecular weights (> $10^8$ Da). However, a ν < 0.33 and a negative α suggest an increasing sphere density with increasing molecular weight. Thus, instead of three conformation areas, a continuous transition may be assumed where a random coil-like structure transforms into a spherical polymer whose density increases with increasing molecular weight.

The reason for these conformational changes can be attributed to a higher density of intramolecular interactions at higher molecular weights. As the chain length increases, the number of intersegmental contacts and therefore, the number intramolecular interactions per chain segment increases. This causes a restriction of movement of the individual fructose segments in the molecule and a restricted rotation around the glycosidic bonds (Kitamura et al., 1994). Benigar et al. (2014) also proposed that intermolecular attractions become stronger with increased molecular weight due to greater entanglement probabilities. Another possible explanation for the structural change to a more compact molecular structure could be an increasing number of branches with increasing molecular weight. Detailed branching analyses of the fractionated levan samples obtained in the present study could hence help to support the latter theory. However, previous $^{13}$C-NMR measurements of two non-fractionated *G. albidus* levans exhibiting distinctly different molecular weight revealed a linear structure with no or at most very few (non-detectable) branches for both levans (Jakob, 2014).

Considering the structure of the fructose molecule, the attractive intramolecular interactions are most likely based on hydrogen bonds. Stivala and Bahary (1978) detected an increase in levan intrinsic viscosity and gyration radius in solution by adding urea and increasing the temperature. Both actions interfere with the formation of the intramolecular hydrogen bonds causing the polymer to swell. Another indication of more pronounced hydrogen bonds with increasing polymer size is the molecular weight dependent Huggins coefficient at high molecular weights. At a certain molecular weight an increased level of inter- and intramolecular hydrogen bonds alters the solubility and the macromolecular conformation of levan. Because of these changes, which are progressively pronounced with the molecular weight, the Huggins coefficient increases. In comparison to other uncharged polysaccharides such as dextran, the low intrinsic viscosity and compact molecular structure of levan might be due the D-fructofuranose ring having greater flexibility than the D-glycopyranose ring (Arvidson et al., 2006). This idea is reinforced by both β-2,6 linked levan and the β-2,1 linked fructose polymer inulin having a low intrinsic viscosity and a compact molecular structure. The similarity of the two polymers is also proved by exhibiting similar conformation changes in the same molecular weight range (Kitamura et al., 1994; Wolff et al., 2000). However, no decrease in intrinsic viscosity of inulin at high molecular weights has yet been reported.

## 5. Conclusions

The present study demonstrates the transformation of *G. albidus* levan from a compact random coil structure to a dense sphere with increasing molecular weight. Moreover, the structure of the spherical levan molecule became more and more compact with increasing molecular weight indicated by a decline in intrinsic viscosity with increasing molecular weight ($M > 10^8$ Da). The structure of a polysaccharide in solution determines its techno-functional properties. Therefore, the findings contribute to establishing the structure-function relationship of levan. Moreover, the results contribute to the mechanistic understanding for the potential use of levan in food, cosmetic and other industrial applications. However, the performance of levan in more complex food formulations is largely unknown. Therefore, the characterisation of levan-protein mixtures in dilute and concentrated systems with respect to molecular interactions, phase behaviour and macromolecular structure formations should be investigated in further studies. For this purpose, the second cross virial coefficient of levan and common food proteins can be characterized using membrane-osmometry to describe the molecular interactions of a binary system. Moreover, the impact of high molecular levan on rheological properties can be investigated in protein-rich systems to establish levan as a functional hydrocolloid of industrial relevance.

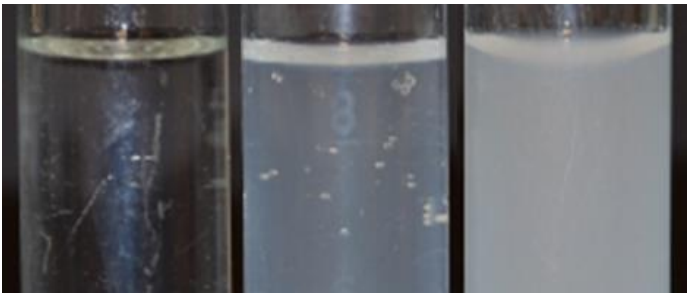

Figure 1: Levan solutions (8 g/L) with increasing polymer size (left to right).

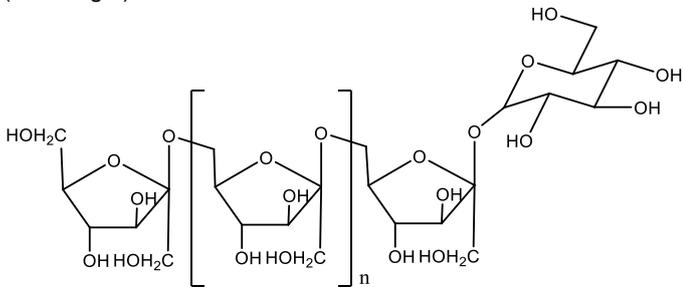

Figure 2: Structure of levan.

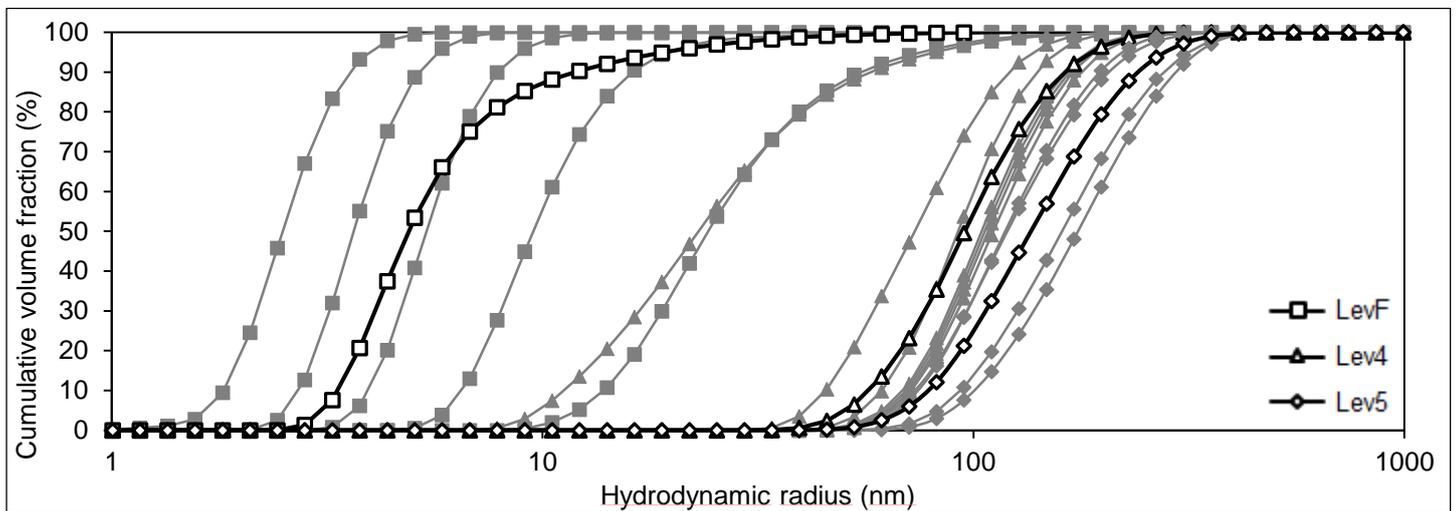

Figure 3: Cumulative distribution of hydrodynamic radius form DLS of levan before (black, unfilled symbols) and after (grey, filled symbols) ethanol fractionation. LevF (squares), Lev4, (triangles), Lev5 (diamonds).

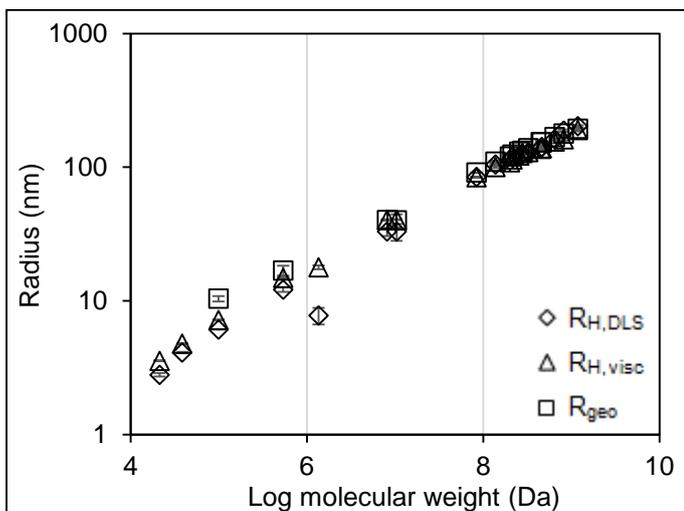

Figure 4: Dependence of geometric Radius $R_{geo}$ (squares), hydrodynamic radius $R_{H,DLS}$ from DLS (diamonds) and hydrodynamic radius $R_{H,visc}$ from viscometry (triangles) on levan molecular weight.

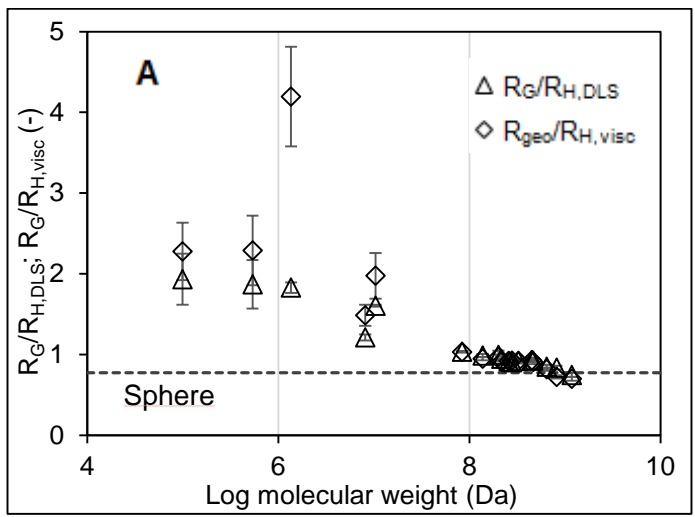
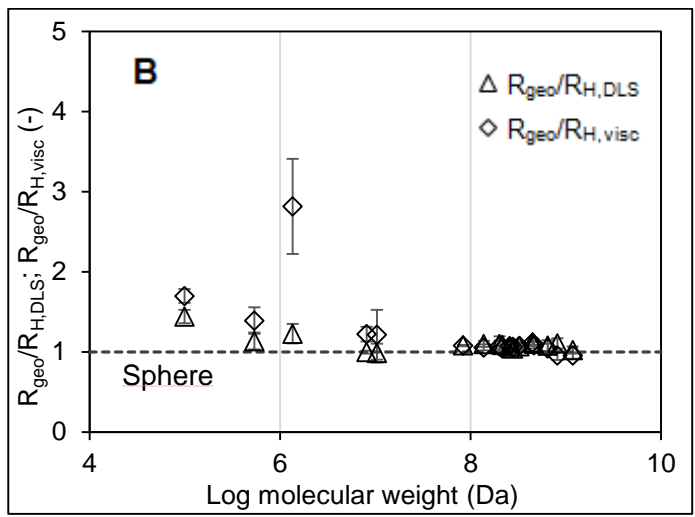

Figure 5 A: Dependence of $R_G/R_{H,DLS}$ (diamonds) and $R_G/R_{H,visc}$ (triangle) on levan molecular weight. B: Dependence of $R_{geo}/R_{H,DLS}$ (diamonds) and $R_{geo}/R_{H,visc}$ (triangle) on levan molecular weight. The dotted line at 0.774 (A) and 1.0 (B) represents the theoretical value for a compact sphere.

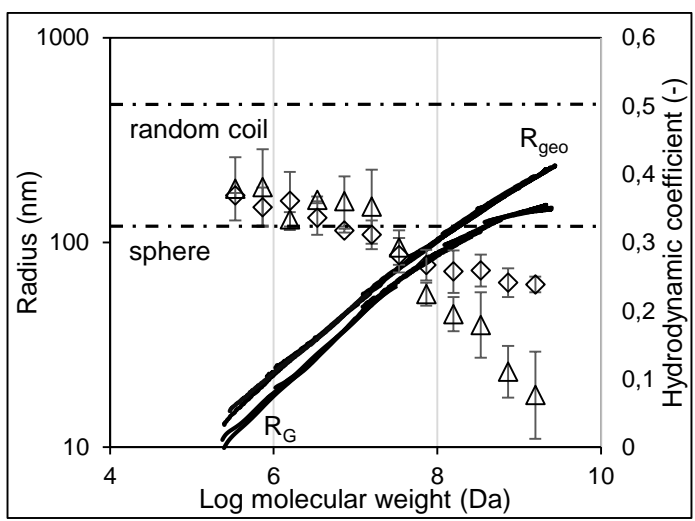

Figure 6: Dependence of geometric radius $R_{geo}$, radius of gyration $R_G$ (lines, right axis) and hydrodynamic coefficients (triangles $v_G$; diamonds $v_{geo}$, left axis) on levan molecular weight. The lines at 0.5 and 0.33 represents the theoretical hydrodynamic coefficients of a random coil at theta conditions and a compact sphere respectively.

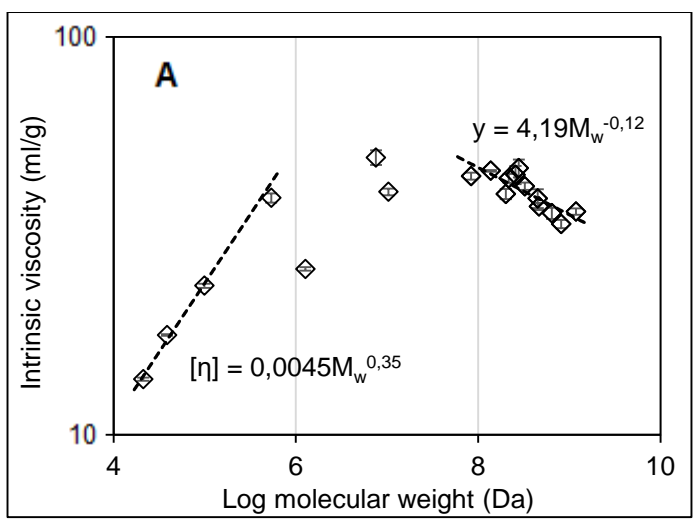
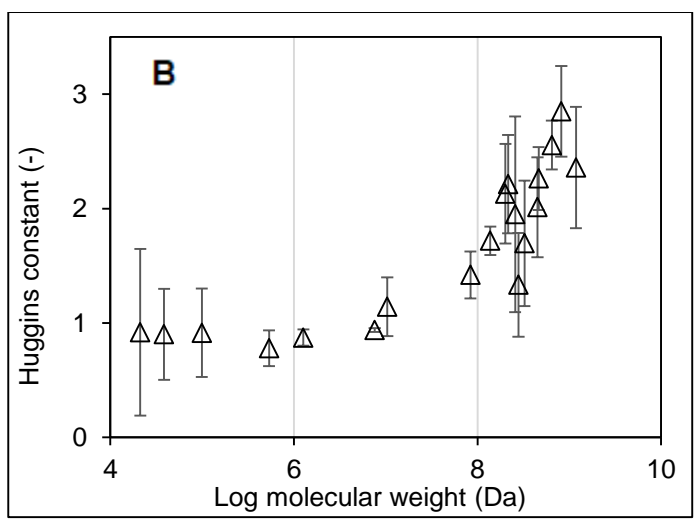

Figure 7 A: Mark-Houwink-Sakurada Plot. B: Dependence of Huggins constant on levan molecular weight.

## Supporting Information

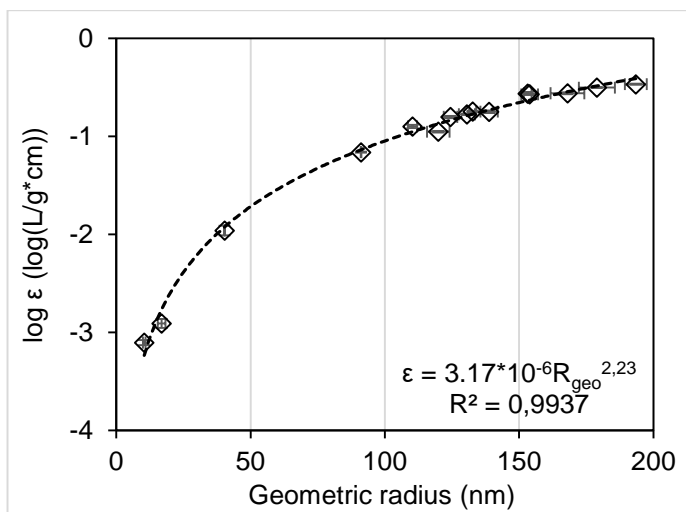

Figure S8: Dependence of specific extinction coefficient on geometric radius for levan.

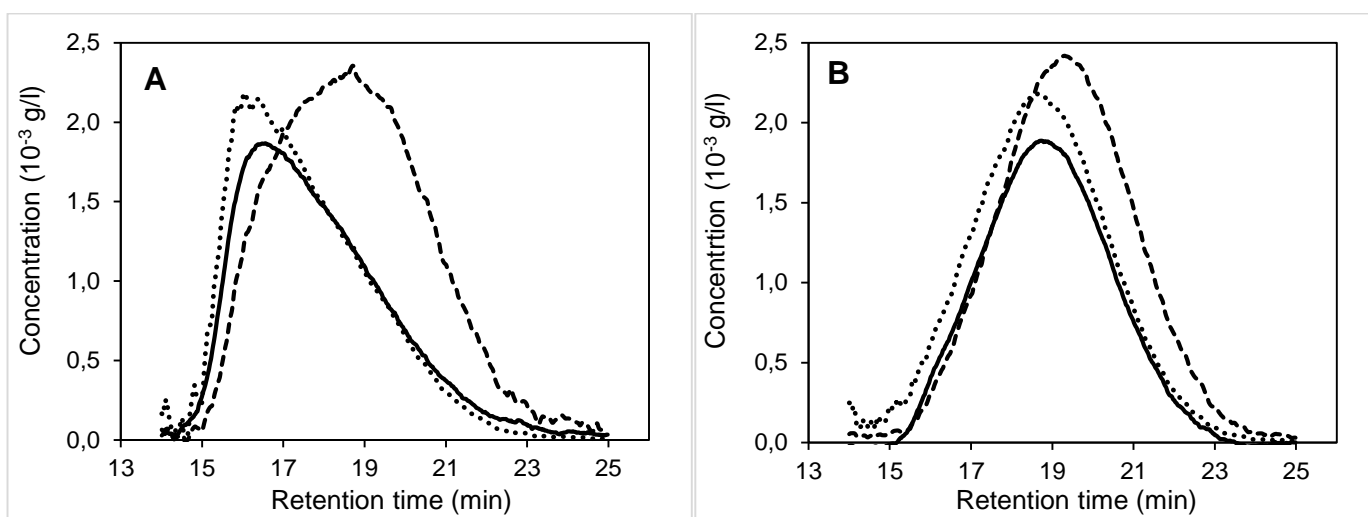

Figure S9: Polymer concentration as a function of the retention time during the aF4-MALLS measurement. Solid line RI detection, dashed line UV detection, dotted line corrected UV detection. A Lev4 F6, B Lev4 F5

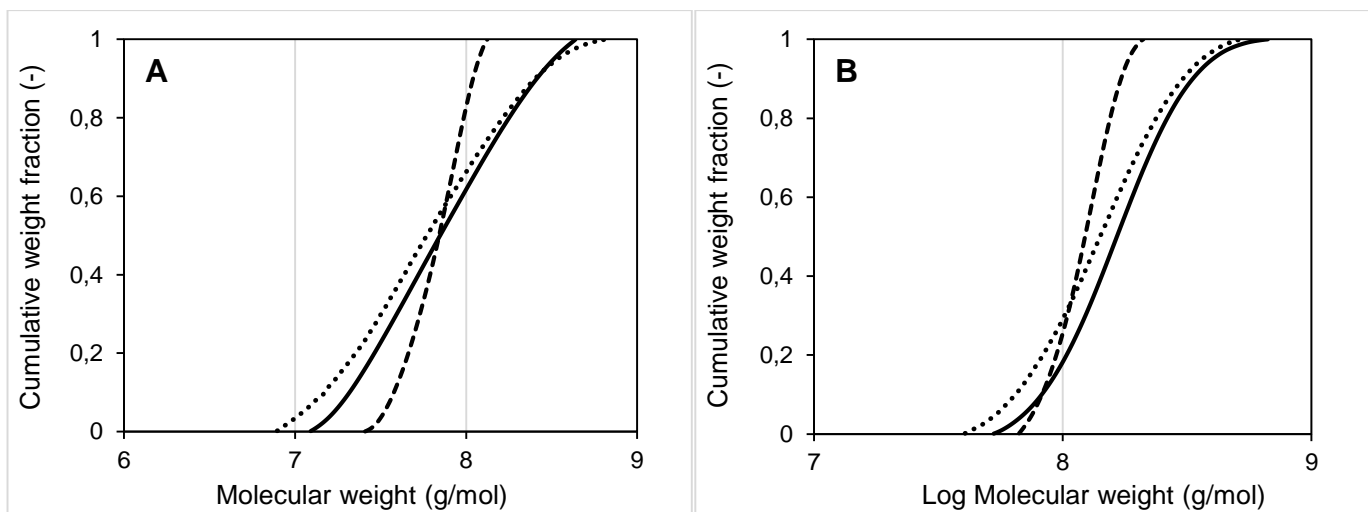

Figure S10: Cumulative molecular weight distribution of Lev4 F6 (A) and Lev4 F5 (B). Solid line RI detection, dashed line UV detection, dotted line corrected UV detection.

Table S1: Molecular Sizes of levan and its fractions

| | Fraction | $M_W$ (Da) | $R_{geo}$ (nm) | $R_G$ (nm) | PDI (-) | $R_{H,DLS}$ (nm) | $R_{H,visc}$ (nm) | $[\eta]$ mL/g |
|---|---|---|---|---|---|---|---|---|
| LevF | | $1.4*10^6$ | 21.9 | 33 | 19.9 | 8 | 18 | 26 |
| | 1 | $8.1*10^6$ | 40 | 48 | 8.1 | 33 | 40 | 50 |
| | 2 | $5.4*10^5$ | 17 | 23 | 2.9 | 12 | 15 | 39 |
| | 3 | $1.0*10^5$ | 10 | 11 | 2.2 | 6 | 7 | 24 |
| | 4 | $3.9*10^4$ | - | - | 2.1 | 4 | 5 | 18 |
| | 5 | $2.1*10^4$ | - | - | 2 | 3 | 4 | 14 |
| Lev4 | | $2.0*10^8$ | 120 | 108 | 1,5 | 112 | 109 | 40 |
| | 1 | $3.3*10^8$ | 139 | 118 | 1.3 | 128 | 130 | 42 |
| | 2 | $2.8*10^8$ | 133 | 115 | 1.2 | 124 | 128 | 47 |
| | 3 | $2.6*10^8$ | 131 | 111 | 1.2 | 122 | 122 | 45 |
| | 4 | $2.2*10^8$ | 124 | 109 | 1.2 | 119 | 115 | 44 |
| | 5 | $1.4*10^8$ | 110 | 99 | 1.4 | 105 | 100 | 46 |
| | 6 | $8.4*10^7$ | 91 | 87 | 2.3 | 85 | 84 | 45 |
| | 7 | $1.0*10^7$ | 40 | 65 | 31.1 | 33 | 41 | 41 |
| Lev5 | | $6.5*10^8$ | 168 | 131 | 1,7 | 161 | 154 | 36 |
| | 1 | $1.2*10^9$ | 193 | 139 | 1.2 | 202 | 190 | 36 |
| | 2 | $8.2*10^8$ | 179 | 136 | 1.4 | 188 | 164 | 34 |
| | 3 | $4.6*10^8$ | 153 | 128 | 1.5 | 142 | 140 | 37 |
| | 4 | $4.5*10^8$ | 154 | 129 | 1.5 | 138 | 141 | 39 |